# Radio sky and the right to observe it

Sergei Gulyaev & Paul Banks

*Institute for Radio Astronomy and Space Research, Auckland University of Technology (AUT University), Auckland, New Zealand*

**Abstract**. It was decided in May 2012 that the Square Kilometre Array (SKA) will be built in Africa and Australia – two Southern Hemisphere continents. Here we discuss the plan for SKA design and construction, and how New Zealand radio astronomers can participate in this project and contribute to astronomy and astrophysics research. Geodesy and the study of tectonic plate motion is another important area of research for New Zealand radio astronomy to contribute to. As New Zealand is located at the boundary between two colliding tectonic plates (Australian and Pacific) and most of geological activity in New Zealand originates from their motion, it is important to monitor the relative plate motion with high precision using both GPS and radio astronomical techniques. We discuss radio frequency interference (RFI) as a limiting factor for radio astronomy, and provide results of RFI measurements in different locations in New Zealand.

## 1. Radio astronomy and SKA

The Square Kilometre Array (SKA) is destined to become the largest and most powerful radio telescope ever built. With a collecting area of a square kilometre, it will be 50 times more sensitive than any existing radio telescope. This will allow it to survey the Universe at radio wavelengths 10,000 times faster than any existing instrument, opening the door to amazing new discoveries. The SKA will be capable of addressing some of the most fundamental questions in physics and astronomy: What is dark energy which makes up 73 percent of the Universe? Was Einstein right about gravity or is there a deeper truth waiting to be discovered? How did galaxies form and how do they change over time? Are we alone?

Fig. 1 shows schematically the sizes (baselines) for four major components of SKA and their frequency ranges. According to the decision made by the SKA Board in May 2012, three thousand 15-m dishes (marked as "Dish-Hi" in Fig. 1) will be built in Africa spreading from the Republic of South Africa to Madagascar in the east and up to Ghana towards the north of the continent. Both the low-frequency aperture array of tiles (shown as "AA-Lo" in Fig. 1) and a 100-dish array ("Dish-Lo" in Fig. 1)

– a revolutionary "survey telescope" – will be built in Western Australia. The SKA construction will be split into two periods/phases, SKA-1 (first 10% of all SKA) and SKA-2 (the rest of SKA). Both SKA-1 and SKA-2 phases will require 4-5 years each. It is possible that one more astronomical instrument – a dense aperture array ("AA-Hi") – will be built in Africa in addition to 3000 dishes as a part of the SKA-2 construction phase.

Nine countries form and fund the SKA Organisation: Australia, Canada, China, Italy, Netherlands, New Zealand, South Africa, Sweden, and the United Kingdom; India is an associate member (2012). A 2-year SKA pre-construction phase has started, which will provide a detailed design of all elements of the first phase, SKA-1. Many universities and industries around the world have formed consortia, each consortium being responsible for the development of one of twelve SKA "work packages", such as Dishes, Low-frequency Aperture Array, Signal and Data Transport, Central Signal Processor, Science Data Processor, Telescope Manager, Synchronisation and Timing, Power, Site and Infrastructure, and so on. They will have to solve problems of Exa-scale computing and Petabit/second connectivity and minimize power consumption – grand challenges for humankind.

This great work started much earlier, when an international group of experts launched the Prep-SKA project, and when construction of SKA precursor telescopes started both in Africa (MeerKAT) and in Western Australia – the Murchison Widefield Array (MWA) and the Australia SKA Pathfinder (ASKAP). According to the SKA master-plan, ASKAP will eventually consist of 100 dishes working as one giant telescope. Each dish of this "survey telescope" will be equipped with new technology developed in Australia called the Phased Array Feed. PAF is a radio analogue of the CCD matrix which revolutionized optical astronomy (see, for example, the report by Fisher, 2010). This work in Australia is led by John O'Sullivan, a co-inventor of Wi-Fi.

New Zealand, like any other country in the world, will certainly benefit from the SKA through new technologies, engineering inventions, astronomical discoveries, the understanding of dark matter and dark energy and so on. No part of the SKA will be built on New Zealand land, but New Zealand academia and industries will be involved in the design work on the SKA and in research projects with the pathfinder radio telescopes, ASKAP and MWA. We want to be a part of the exciting process of SKA development.

Recently, AUT University's 12-m radio telescope in Warkworth (Gulyaev and Natusch, 2010) was equipped with new "L-band" feed that makes it sensitive to radio waves around 21 cm and 18 cm wavelengths – the interval called the "water gap" because it corresponds to the radiation of neutral hydrogen (H) and the hydroxyl molecule maser (OH). This important development makes the New Zealand radio telescope compatible with ASKAP, which will operate mainly in L-band.

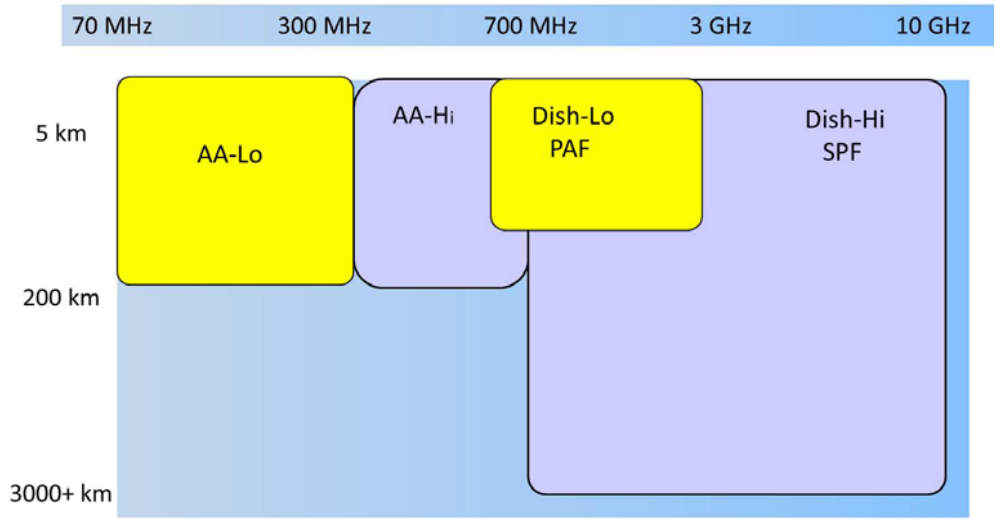

Fig. 1. Baseline vs. Frequency for four major components of the SKA. The components to be built in Australia are marked in yellow. PAF is the phased array feed, SPF is the single pixel feed.
After R. Schilizzi

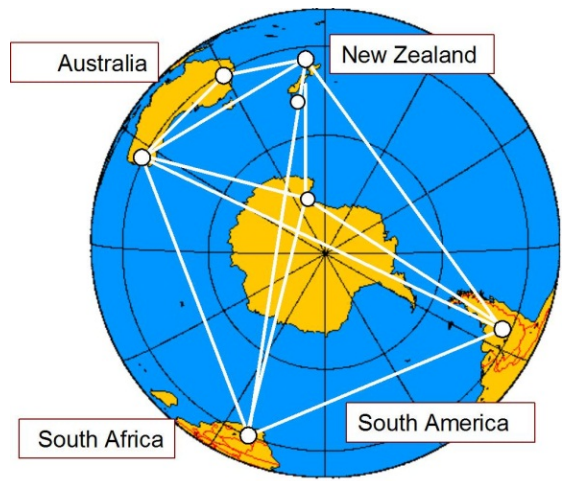

Fig. 2. Geographical distribution of SKA nations – Australia, New Zealand, South Africa. Major baselines are shown. Locations of existing antennas in South America and possible antennas in New Zealand's South Island and in Scott Base (Antarctica) are indicated.
Projection: polar orthographic

After 3000 SKA dishes are built in Africa, all Southern Hemisphere radio telescopes (including New Zealand's) will be able to join together in an unprecedented, extremely sensitive, array of telescopes of the size of the Earth. We also look forward to joint VLBI (very long baseline interferometry) observations in L-band with a mighty 100-dish ASKAP.

## 2. Radio astronomy and tectonic plate motion

Geodesy is another important area of science where New Zealand radio astronomy is involved (Blick & Gulyaev, 2009). By observing thousands of quasars, the most remote objects in the Universe, radio astronomers create a reference frame in which all possible motions of the Earth and its tectonic plates can be studied. This reference frame is called the International Celestial Reference Frame and it is the basis of the International Terrestrial Reference Frame, which is in turn the basis for all geographic coordinate systems and maps. While the SKA contributes to fundamental science, geodesy is an extremely important applied science. One of applications of geodesy is, for example, the study of tectonic plate motions responsible for different types of geological activity, including volcanic eruptions and earthquakes. International The VLBI Service for Geodesy and Astrometry (IVS) is the organisation that coordinates geodetic radio astronomical observations around the globe.

New Zealand's radio telescope in Warkworth is one of many world radio telescopes that contribute to the IVS network and activities. Located in the eastern part of the Australian tectonic plate, it (together with a number of Australian radio telescopes) provides important information about the status of this plate and its internal motions. A very accurate tie (to 1 mm accuracy) between this radio telescope and a network of GPS stations located on the New Zealand's North Island, allows the use of *both* powerful geodetic techniques – VLBI and GPS – to jointly monitor the Australian tectonic plate deformation and to support the New Zealand national coordinate system, the New Zealand Geodetic Datum.

The power of the VLBI technique when used for geodesy, lies in providing direct measurements of baselines, which are the distances between radio telescopes. The use of the technique for earthquake research reveals the most when the telescopes are located on different tectonic plates. Monitoring baselines and their time derivatives – velocity, acceleration, and even higher derivatives – provides information which, we believe, will allow scientists to develop methods for earthquake prediction.

Currently New Zealand has one geodetic radio telescope in the North Island at Warkworth, which is on the Australian tectonic plate. For New Zealand scientists to be able to use the VLBI technique for monitoring the relative motion of the Australian and Pacific tectonic plates, a radio telescope on the Pacific tectonic plate in the South Island is required. Given that major earthquakes have devastated Christchurch over the last few years, and given that "the big one" expected across the main divide of the South Island is yet to occur, we see it as essential that a radio telescope is constructed on the Pacific plate as soon as possible in order to gain the desperately needed information and insight offered by this technique. The personnel, and even much of the equipment required, are already available on both tectonic plates, and international expertise is readily available to assist with any telescope construction. The most pressing need is for immediate construction of a single southern telescope which would provide a VLBI link to Warkworth.

## 3. Radio Frequency Interference

The location of the "Pacific plate radio telescope" in the South Island of New Zealand should be determined from a number of criteria. For example, it should be located far enough from the boundary between two tectonic plates to avoid minor earthquakes that may affect positioning results. In this regard, east Otago and Southland are good locations for a geodetic radio telescope. Existing infrastructure is also a consideration. The most important criterion is the requirement for a low level of radio frequency interference (RFI).

RFI is the analogue of light pollution in optical astronomy. To protect radio astronomical observations from RFI, special frequency bands are allocated and reserved for radio astronomical use only. For example, the band 1400-1427 MHz is allocated to protect radio observations in the neutral hydrogen (HI) spectral line (21-cm spectral line); an allocated band at 1660-1670 MHz protects observations of the hydroxyl molecule (OH) maser (18-cm spectral lines), and so on (Fig. 3). No man-made transmitters are allowed to radiate at the radio astronomy allocated frequencies. The International Telecommunication Union (ITU) is the radio frequency spectrum controlling authority worldwide. The protection criteria for the radio astronomical measurements are presented in the ITU's Recommendation ITU-R RA.769-2 – Protection criteria used for radio astronomical measurements (2003).

Allocation of special bands and well-developed mechanisms for their protection are very important achievements of the radio astronomical community; they are not sufficient, however, for modern extragalactic radio astronomy. Consider, for example, the HI spectral line at 21-cm wavelength. It is formed as the result of transitions between sublevels of the hyperfine structure of the ground state of atomic hydrogen. The laboratory frequency of this transition is 1420.4 MHz, which means that it is closer to the upper limit of the allocated 27-MHz-wide protection band, than to its lower limit.

Using a formula for the redshift, $1 + z = \nu_{em}/\nu_{obs}$, where $\nu_{em}$ is the frequency emitted by the radio source with the redshift $z$, and $\nu_{obs}$ is the frequency observed with the ground based radio telescope, we find that the allocated band allows us to observe radio sources in the range of redshifts from $z = -0.005$ to $z = 0.014$. This interval is sufficient for studying the Milky Way and relatively close galaxies, but it is not sufficient for high-redshift radio sources. Even for radio sources with the redshift $z = 0.1$, the lower limit of the ITU allocated band should be shifted down to 1280 MHz (instead of the current 1400 MHz). Due to the redshift, the observed frequencies for radio sources with redshifts $0 < z < 1$ are in the range of 710.2 to 1420.4 MHz (Fig. 3).

The SKA will be designed to study radio galaxies with a redshift greater than $z = 1$ or even greater than $z = 2$ ($\nu_{obs}$ =473 MHz). In fact, the highest confirmed spectroscopic redshift of a galaxy is $z = 8.6$ (Lehnert et al., 2010). Therefore, the ITU allocated frequencies serve only those areas of radio astronomy that study our Galaxy and its neighbourhood. As the boundaries of scientific interests extend to very remote (high-redshift) objects in the Universe, the frequencies of radio astronomical interest go far beyond the ITU allocated boundaries into parts of the radio spectrum used by other active users. That makes it important for radio astronomers to look for effective means of RFI protection and mitigation. These include methods based on the use of additional hardware (e.g. radio frequency filters to filter out the RFI signals) as well as more recent methods of software mitigation of RFI. A natural means of RFI protection remains a careful choice of location for radio telescopes. A survey for a suitable low-RFI location – a radio quiet zone (RQZ) – needs to be undertaken before the construction of a new radio telescope.

In New Zealand an extensive RFI survey for selected locations was undertaken by AUT University's Institute for Radio Astronomy and Space Research in 2008-2009 (Banks, 2009). This work was sponsored by the New Zealand Ministry for Economic Development and conducted in collaboration with CSIRO Astronomy and Space Science – CASS (Australia). Resulting radio spectra below 1500 MHz for several locations in New Zealand are shown in Fig. 4.

While this frequency band may be of only limited interest to astronomers, the charts contain aspects that do make them interesting. Firstly, they show a clear difference in radio traffic levels between city and rural locations similar to light pollution variance; and secondly, they show the main interferers. The lower frequency end of the radio spectrum normally contains the thumpers, or most powerful transmissions that are likely to overload the amplifiers of a radio telescope, creating ghost signals, and must therefore be filtered off. The problem with filtering is that signals of interest are attenuated, hence the need for surveys to see what RFI exists at a site in the first place. More importantly, most transmissions occurring in this band are below 1000 MHz. The charts provide received transmission power amplitudes across frequency bands of interest, averaged, in these cases, over a night. They provide maximum, mean and minimum RFI levels, but the code can be altered to provide 90 percentile and median levels determined as required for SKA site evaluation by Ambrosini et al. (2003).

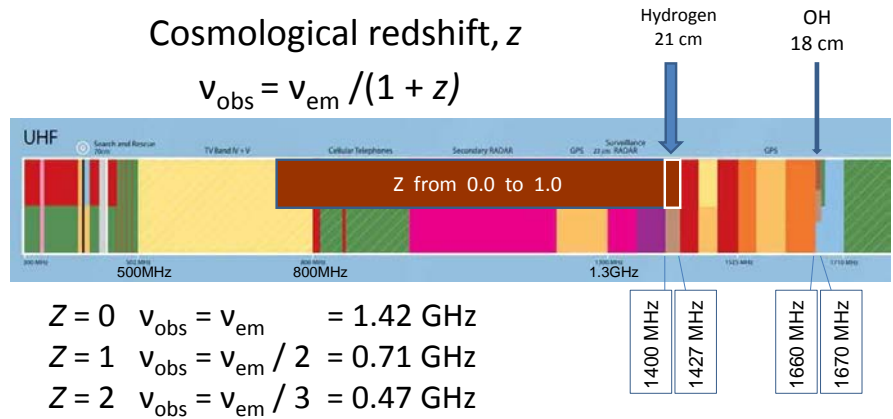

$Z = 0$   $v_{obs} = v_{em}$        $= 1.42$ GHz
$Z = 1$   $v_{obs} = v_{em} / 2$ $= 0.71$ GHz
$Z = 2$   $v_{obs} = v_{em} / 3$ $= 0.47$ GHz

Fig. 3. New Zealand radio spectrum allocation and radio astronomy bands for neutral hydrogen (21 cm band) and for hydroxyl molecule OH (18 cm band). The role of the redshift is illustrated; frequencies of the hydrogen line for SKA objects with $0 < z < 1$ are marked.

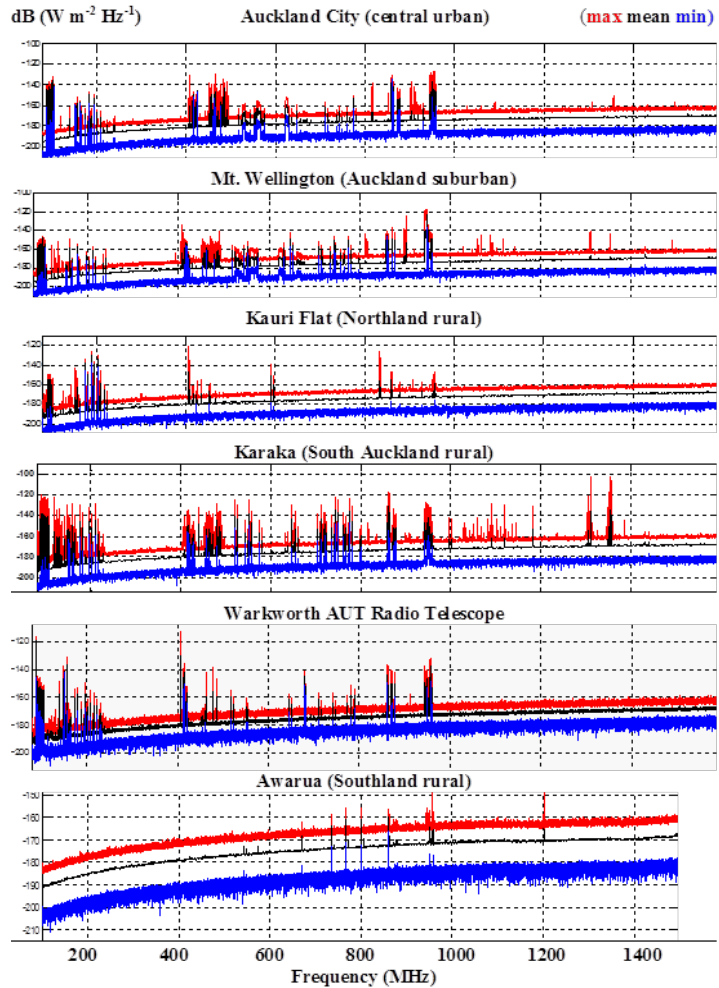

Fig. 4. RFI radio spectra (power in dB vs. frequency MHz) as measured for several locations in New Zealand. Maximum, mean and minimum RFI levels are shown in red, black and blue respectively (Banks, 2009).

## 4. Conclusion

In conclusion, as the proof of pudding is in eating, the successful work of the Warkworth radio astronomical observatory and RFI measurements demonstrate that New Zealand has excellent radio-quiet locations for radio astronomical observations. There is a need for a radio telescope in the South Island to ensure that the significant contribution of VLBI to the study of tectonic motions responsible for catastrophic earthquakes is able to be made within New Zealand. New Zealand scientists will contribute to the SKA in its pre-construction and construction phases. On its completion, New Zealand will be a part of the most powerful astronomical array linking together radio telescopes of the Southern Hemisphere.